\documentclass[aps,twocolumn, superscriptaddress,showpacs]{revtex4}

\usepackage{graphicx}
\usepackage{amsmath}
\usepackage[utf8]{inputenc}

\begin{document}

\title{Charge and spin transport in a metal-semiconductor heterostructure with double Schottky barriers}

\author{S. Wolski}\email{wolski@prz.edu.pl}
\affiliation{Department of Physics,
Rzesz\'ow University of Technology, Al.~Powsta\'nc\'ow Warszawy 6,
35-959 Rzesz\'ow, Poland}

\author{C. Jasiukiewicz}
\affiliation{Department of Physics,
Rzesz\'ow University of Technology, Al.~Powsta\'nc\'ow Warszawy 6,
35-959 Rzesz\'ow, Poland}

\author{V.~K.~Dugaev}
\affiliation{Department of Physics, Rzesz\'ow University of Technology,
Al.~Powsta\'nc\'ow Warszawy 6, 35-959 Rzesz\'ow, Poland}
\affiliation{Departamento de F\'isica and CFIF, Instituto Superior T\'ecnico,
Universidade de Lisboa, Av. Rovisco Pais, 1049-001 Lisbon, Portugal}

\author{J. Barna\'s}
\affiliation{Faculty of Physics, Adam Mickiewicz University, ul. Umultowska 85, 61-614 Pozna\'n, Poland}

\author{T. Slobodskyy}
\affiliation{Institute for Applied Physics, University of Hamburg,\\ Jungiusstraße 11, 20355 Hamburg, Germany}

\author{W. Hansen}
\affiliation{Institute for Applied Physics, University of Hamburg,\\ Jungiusstraße 11, 20355 Hamburg, Germany}

\begin{abstract}
Taking into account the available experimental results, we model the electronic
properties and current-voltage characteristics of a ferromagnet-semiconductor junction.
The Fe/GaAs interface is considered as a Fe/($i$-GaAs)/$n^{+}$-GaAs/$n$-GaAs
multilayer structure with the Schottky barrier.
We also calculate numerically the current-voltage characteristics of a double-Schottky-barrier structure
Fe/GaAs/Fe, which are in agreement with available experimental data. For this structure, we
have estimated the spin current in the GaAs layer, which characterizes spin injection from the ferromagnet
to the semiconductor.
\end{abstract}

\date{\today }
\pacs{73.30.+y, 73.40.-c, 75.76.+j}
\maketitle

{\it Introduction.}
One of the key issues in spintronics is the problem of efficient spin
injection and spin manipulation in semiconductors \cite{zutic04}. It is well known that
when spin-polarized electrons are directly transmitted from a ferromagnetic metal into a semiconductor,
efficiency of such spin injection is usually very low
\cite{lee99,hammar99} because of a large difference in the conductivities of the metal and semiconductor
\cite{schmidt00,khaetskii05}. However,
a tunneling barrier between the metal and semiconductor can help in maintaining the
spin efficiency relatively high \cite{rashba00,smith01}. Accordingly, the Schottky barrier at the metal-semiconductor
interface \cite{sze} attracted a lot of attention since it can act as a natural tunnel barrier
separating the semiconductor and the ferromagnetic metal.

The Schottky barrier \cite{rhoderick82,sze,tung14} with optimal parameters for spin injection can appear
only at certain conditions related to the physical processes
at the metal-semiconductor heterojunction. All the parameters of the Schottky barrier, such as its
height and width and the energy profile, mostly depend on the distribution of dopants in the
semiconductor near the interface.
When the external voltage is applied to the barrier, the accumulated electric charge related
to redistribution of electrons and holes strongly affects the energy profile. Apart from this,
 a nonequilibrium spin accumulation appears near the
ferromagnet-semiconductor interface,
which affects transport of electrons and holes in different spin channels.

Our main objective is to find optimal regimes of charge and spin transport through the Schottky
barrier, where spin injection to the semiconductor is efficient. To do this we
simulate physical processes in the junction, taking into account modification of the electronic
energy structure near the interface as well as the spin and charge accumulation in a nonequilibrium situation.
As a result of the numerical simulations, we
obtained the current-voltage characteristics
of the Fe/GaAs/Fe junction with a double Schottky barrier. These results are in satisfactory agreement
with the  $I-V$ characteristics obtained experimentally on the Fe/$n^{+}$-GaAs/$n$-GaAs/$n^{+}$-GaAs/Fe
heterostructure  which works as the double Schottky barrier.

\vskip0.5cm
{\it Model.}
We describe the charge and spin transport in ferromagnet-semiconductor 
structures in terms of a semiclassical model, with thermally activated 
electrons and holes near the metal-semiconductor interface. This model 
properly describes transport properties at room temperatures, where the 
thermally
activated conductivity is much larger than the under-barrier tunneling.
In the semiclassical approximation we have to calculate spin-resolved 
profiles for
the electrostatic potential, spin-polarized electron and hole densities, and 
the chemical potential.
The key point of the simulations is self-consistency, which is especially 
important for a large
deviation of the system from equilibrium.

\begin{figure}[htb]
\hspace*{0 cm}
\includegraphics[width=0.8\linewidth]{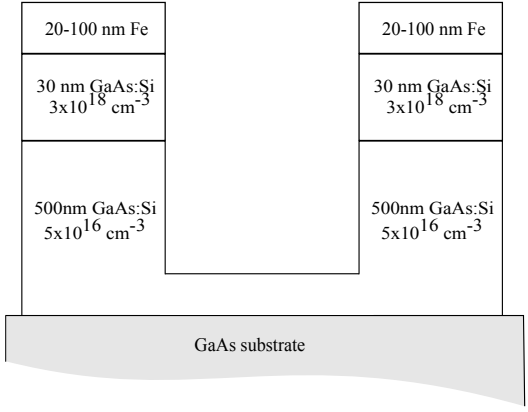}
\vspace*{-0.2cm}
\caption{Schematic view of the Fe/GaAs/Fe heterostructure with two Schottky junctions.
}
\label{fig:schottky-schema}
\end{figure}

The heterostructure under consideration is shown schematically in Fig.~\ref{fig:schottky-schema}.
The doping concentration is $N_D = 5\times 10^{16}$~cm$^{-3}$.
For lower doping concentration, the conductance is significantly reduced.
The 30~nm $n^{+}$-GaAs layer of enhanced donor concentration assures efficient spin injection trough the barrier
\cite{hu11,crooker09}.

The Poisson equation, which accounts for space distribution of the total charge, determines the
electrostatic potential $\phi (x)$,
\begin{equation}
-\nabla ^2\phi =4\pi e\, (n+N_a^{-} -N_d^{+} -p),
\label{eq:poisson}
\end{equation}
where the electron and hole concentrations, $n=n(x)=n_\uparrow (x)+n_\downarrow (x)$ and
$p=p(x)=p_\uparrow (x)+p_\downarrow (x)$, respectively,
are calculated from the corresponding spin components determined by spin-resolved
chemical potentials $\mu _\sigma (x)$, while $N_a^-(x)$ and $N_d^+(x)$ are the densities of
ionized acceptors and donors.
The electric current in each spin channel is related to the gradient of the corresponding chemical potential,
$j_{n\sigma }=\mu _n n_\sigma \, \nabla \mu _\sigma $ for electrons and
$j_{p\sigma }=\mu _p p_\sigma \nabla \mu _\sigma $  for holes, where $\mu _n$ and $\mu _p$
are the mobilities of electrons and holes.
The difference of chemical potentials in each spin channel at the ends of a sample is determined
by the external voltage $V$.

The energy band diagram in Fig.~2 shows schematically the equilibrium profiles of the conductance ($\varepsilon_c$)
and valence ($\varepsilon_v$) band edges in GaAs in the double junction structure. The curvature of the
band edges is due to the spatial variation of electrostatic potential,
$\varepsilon _{c,v}=\varepsilon _{c0,v0}+e\phi $, where $\varepsilon _{c0,v0}$ describe the band edges in the corrresonding bulk semiconductor.

\begin{figure}[htb]
\vspace*{0.0cm}
\hspace*{0 cm}
\includegraphics[width=1.0\linewidth]{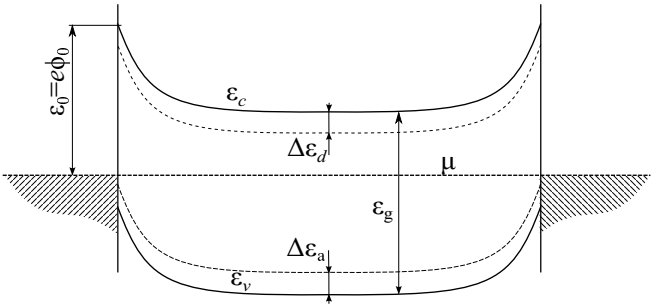}
\vspace*{-0.2cm}
\caption{Energy band diagram of the double metal-semiconductor junction in thermal equilibrium. Here $\mu$
is the equilibrium chemical potential, $\Delta \varepsilon_d$ and $\Delta \varepsilon_a$ describe the donor and acceptor levels, respectively,
$\varepsilon_g$ is the energy gap, while
$\varepsilon_0$ denotes the work function.  }
\label{fig:energy-diagram}
\end{figure}

\begin{figure}[pbt]
\vspace*{-0.5cm}
\hspace*{-0.2cm}
\includegraphics[width=1.05\linewidth]{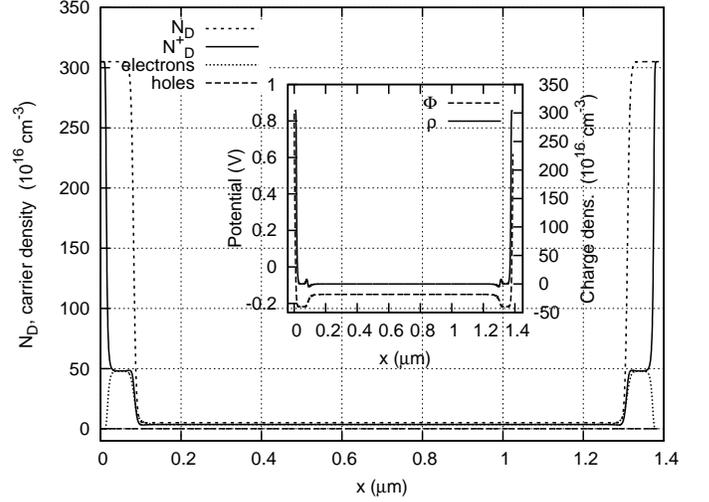}
\caption{Spatial distribution of donors, electrons and holes in
Fe/$n^{+}$-GaAs/$n$-GaAs/$n^+$-GaAs/Fe heterostructure under a constant current. The inset shows
the profiles of the electrostatic potential and the charge density along the sample.}
\label{fig:dopping-profile}
\vspace*{-0.2cm}
\end{figure}

\vskip0.5cm
{\it Results.}
The results of numerical calculations of the distribution of electrons, holes and ionized donors in
the biased Fe/$n^{+}$-GaAs/$n$-GaAs/$n^{+}$-GaAs/Fe  heterostructure are presented in
Fig.~\ref{fig:dopping-profile}. Here we take $N_D=5\times 10^{16}$~cm$^{-3}$ for n-GaAs,  
$N_D^+=3\times 10^{18}$~cm$^{-3}$ for $n^+$-GaAs, and $\varepsilon _0=0.84$~eV for the work function.
A constant current $j=5\times 10^{-6}$A/cm$^2$ flows through the structure due to an external voltage,
see also the inset of
Fig.~\ref{fig:dopping-profile}. One can also note, that variation of the electron density is correlated
with the density of ionized donors $N_d^{+}(x)$ within the entire sample.

\begin{figure}%
\vspace*{-0.5cm}
\hspace*{-0.2cm}
\includegraphics[width=0.9\linewidth]{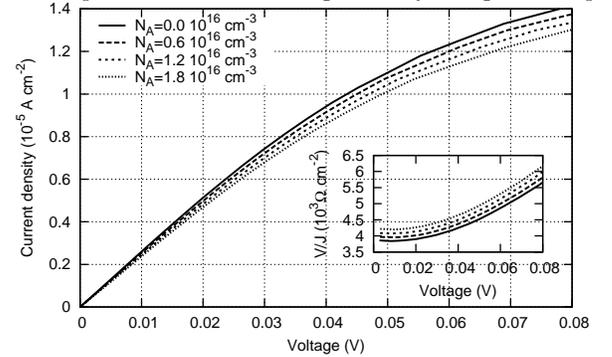}
\caption{Voltage  dependence of electric current for different acceptor density in $n^{+}$-GaAs layer.
The inset shows the dependence of resistivity on applied voltage for different densities of acceptors.}
\label{fig:current-NA}
\vspace*{-0.2cm}
\end{figure}

From detailed analysis of the technology process and  available experimental data on Fe/GaAs/Fe structures \cite{nickiel12}
follows that an important element of the modeling of electric properties of such structures is
an additional thin layer of $i$-GaAs (GaAs strongly doped with Fe), which can be formed in the
vicinity of Fe/GaAs interface \cite{mcinturff96,schieffer06}.
This is due to diffusion of Fe atoms into GaAs, which strongly affect the properties of GaAs
semiconductor making it almost insulating due to the deep-acceptor properties of Fe in GaAs \cite{kleverman83}.
The variation of $I-V$ characteristics with the change of acceptor density in the
$n^{+}$-GaAs layer is presented in Fig.~\ref{fig:current-NA}.

Due to the ferromagnetic elements of the junction, a spin accumulation may appear in the nonequilibrium state.
This accumulation can be included in the numerical simulations in terms of the approach developed in Refs.~\cite{son87} and
\cite{schmidt00}.
In this approach, the spatial variation of the chemical potential is described by the following equation
\vspace*{-0.3cm}
\begin{equation}
\vspace*{-0.3cm}
\frac{d\mu_{\uparrow, \downarrow}}{dx}
= -\frac{e}{\sigma_{\uparrow, \downarrow}}j_{\uparrow, \downarrow},
\label{eq:spin-acc}
\end{equation}
with linear relations between the spin-resolved and total conductivities and currents
\vspace*{-0.3cm}
\begin{align}
\vspace*{-0.5cm}
\label{eq:spin-cond-currents}
{\sigma}_{\uparrow} = \alpha \sigma, \;\;
{\sigma}_{\downarrow} = (1-\alpha) \sigma , \\
{j}_{\uparrow} = \beta j, \;\;
{j}_{\downarrow} = (1-\beta) j, \nonumber
\end{align}
where the coefficients $\alpha $ and $\beta $ describe specific properties of the junction.
\begin{figure}[hbt]
\vspace*{-0.5cm}
\hspace*{-0.2cm}
\includegraphics[width=0.75\linewidth]{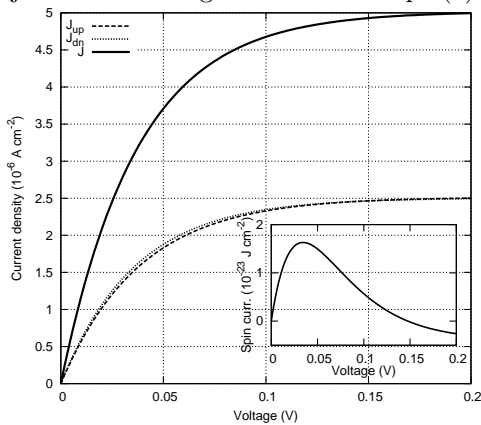}
\vspace*{-0.2cm}
\caption{Current-voltage and spin-current-voltage characteristics for $\alpha =0.501$ and
$\beta =0.5$.}
\label{fig:spin-acc}
\end{figure}
Taking into account  Eqs.~\eqref{eq:spin-acc} and \eqref{eq:spin-cond-currents}, one arrives at the following equation for
the difference between electrochemical potentials for the spin-up and spin-down electron channels,
\vspace{-0.35cm}
\begin{equation}
\vspace*{-0.35cm}
\frac{d( {\mu}_{\uparrow} - {\mu}_{\downarrow})}{dx}
= - \frac{\beta - \alpha}{\alpha (\alpha - 1)} \frac{{ej}}{{\sigma}}.
\label{eq:spin_up_dn_acc}
\end{equation}
Having found the  electrochemical potential and current in both spin channels, one can calculate the  spin current density $j_s$  as $j_s=\hbar (j_\uparrow -j_\downarrow )/2e$.

The results of numerical calculation of the charge and spin current densities are presented
in Fig.~\ref{fig:spin-acc}. For simplicity, we assumed there some effective values of the parameters $\alpha$ and $\beta$; 
$\alpha =0.501$ and $\beta =0.5$.
We note that the spin polarization of transferred current strongly depends
on the factors $\alpha $ and $\beta $, which characterize the interfaces.  
In particular, at the
ferromagnetic side of the interface, they are  determined mainly by the diffusion constant and spin-flip
relaxation time \cite{son87}.

{\it Conclusions.}
Using the semiclassical approach we have calculated the current-voltage characteristics and
spin current in Fe/GaAs/Fe double-Schottky-barrier structures. Our results are in satisfactory agreement
with experimental results of Ref.~\cite{nickiel12}. We have found that the electrical properties of
Fe/GaAs strongly depend on the formation of an additional thin insulating $i$-GaAs
layer near the interface due to diffusion of Fe atoms into GaAs. The spin injection through
the interface is mostly determined by the conductive properties of Fe near the interface and
by the spin-flip scattering in this region.

\vspace{0.3cm}
{\it Acknowledgements.}
This work is supported by the National Center of Research and Development in Poland
in frame of EU project Era.Net.Rus "SpinBarrier".

\end{document}